\documentclass[secnumarabic, nobibnotes, nofootinbib]{ptephy_v1}
\usepackage{amsmath,amssymb}
\usepackage{graphicx}
\usepackage{mathrsfs}
\usepackage{url}
\newcommand{\bs}[1]{\ensuremath{\boldsymbol{#1}}}

\newcommand{\dA}{\delta\!\hat{A}}
\newcommand{\dPhi}{\delta\Phi}
\def\dd{\mathrm{d}}

\begin{document}

\title{Parametrized test of parity--violating gravity using GWTC--1 events}
\author[1,*]{Kei~Yamada}
\affil{Department of Physics, Kyoto University, Kyoto 606-8502, Japan}

\author[1,2]{Takahiro~Tanaka}
\affil{Center for Gravitational Physics, Yukawa Institute for Theoretical Physics, Kyoto University, Kyoto 606-8502, Japan
  \email{k.yamada@tap.scphys.kyoto-u.ac.jp}}

\date{\today}

\begin{abstract}
  Parity--violating (PV) gravity has recently attracted interest in several aspects. One of them is the axion--graviton coupling to test the axion--dark matter model. Moreover, by extending Chern--Simons (CS) gravity to include derivatives of a scalar field up to the second order, a more general class of PV gravity theory, which we call the CNCL model, has been proposed~[M. Crisostomi {\it et al.}, Phys. Rev. D, {\bf 97}, 044034 (2018)]. The model can be further extended by including even higher derivatives of the scalar field and/or higher curvature terms.

  In this paper, we discuss the effect of parity violation in the gravitational sector on the propagation of gravitational waves from binary coalescence by introducing a model--independent parametrization of modification. Our parametrization includes the CNCL model as well as CS gravity. The effect of parity violation on the gravitational waveform is maximum when the source binary orientation to our line of sight is edge--on, while the modified waveform reduces to the parity--symmetric one when the source is face--on. We perform a search for the signature of such modification by using the LIGO/Virgo O1/O2 catalog. We find that the catalog data is consistent with general relativity and obtain constraints on parity violation in gravity for various post--Newtonian order modifications for the first time. The obtained constraint on CS gravity is consistent with the results in previous works. On the other hand, the constraint on the CNCL model that we obtain is tighter than the previous results by roughly 7 orders of magnitude.
    
\end{abstract}

\maketitle

\section{Introduction}

% 1. GWs
The direct detection of gravitational waves (GWs) enabled a new test of general relativity (GR) in strong gravity regimes. 
LIGO and Virgo collaborations (LVC) have reported 10 binary black hole (BH) mergers and one neutron star (NS) binary merger in the catalog GWTC--1~\cite{LIGOScientific:2018mvr}. 
Testing GR has been pursued by several authors using these event data and no significant deviation from GR has been detected~\cite{TheLIGOScientific:2016src, Yunes:2016jcc, Abbott:2018lct, LIGOScientific:2019fpa}. 
Although one of the most interesting regimes to investigate the nature of black holes is the ringdown phase~\cite{Brito:2018rfr, Nakano:2018vay}, there are difficulties in establishing concrete modeling of the merger--ringdown waveform in modified gravity theories~\cite{Berti:2018vdi}. 
% Whilst, the inspiral phase can be studied by employing the post--Newtonian (PN) approximation.
The inspiral phase, meanwhile, can be studied by employing the post--Newtonian (PN) approximation.

% 2. testing GR & the PPE framework
A possible approach to prepare the waveform in modified gravity theories is to use the parametrized post--Einsteinian (PPE) framework, in which the so--called PPE parameters are introduced to describe modifications of gravity theory without specifying the origin of modification~\cite{Yunes:2009ke, Chatziioannou:2012rf, Sampson:2013lpa, Loutrel:2014vja, Huwyler:2014gaa}. 
The PPE framework has since been extended to include non--tensorial polarizations~\cite{Chatziioannou:2012rf}, higher--order corrections~\cite{Sampson:2013lpa}, and eccentric binaries~\cite{Loutrel:2014vja}. 
The waveform in the time domain was also discussed in~\cite{Huwyler:2014gaa}. 
The mapping between the PPE parameters and the model parameters in non--GR models of gravity is known for several cases~\cite{Tahura:2018zuq}. 
As for the modification in the inspiral phase of GWs, the PPE parameters have been constrained via GW observations~\cite{LIGOScientific:2019fpa,Yunes:2016jcc}. 
However, the PPE framework does not cover all the viable extensions of gravity. 
One of the possible extensions which go beyond the scope of the PPE framework is to consider modified GW propagation in parity--violating (PV) gravity.

% 3. Gravitational parity violation
In some candidates of the fundamental theory of gravity, such as string theory and loop quantum gravity, the parity violation in gravity is ubiquitous~\cite{Alexander:2009tp}. 
Gravitational parity violation has been studied most extensively in the context of Chern--Simons (CS) gravity as a concrete example (for a review, see Ref.~\cite{Alexander:2009tp}). 
Recently, CS gravity coupled with an axion field has attracted renewed interest in the context of the parametric resonance of GWs~\cite{Yoshida:2017cjl,Chu:2020iil,Jung:2020aem,Fujita:2020iyx}. 
Moreover, there is a variety of PV gravity models other than CS gravity. 
In Ref.~\cite{Crisostomi:2017ugk}, the authors proposed a more general class of PV gravity theory by extending CS gravity to include derivatives of a scalar field up to second order, which we refer to as the CNCL model.
This model has been constrained from the arrival--time difference between GWs and photons for GW170817~\cite{Nishizawa:2018srh} and more recently the authors of Ref.~\cite{Wang:2020pgu,Wang:2020cub} (see also the references therein)  claimed that a more strict constraint is obtained by analyzing LVC GW events.%
\footnote{In Ref.~\cite{Wang:2020pgu,Wang:2020cub}, the authors restricted themselves in the case that the deviations of waveforms from GR is small and expanded their waveform in small quantities from the beginning. 
This may make their constraint outside/marginal of the validity of their assumption. 
Therefore, in this paper we refer the constraint obtained in Ref.~\cite{Nishizawa:2018srh} as the current one.} 
However, the model can be further extended by including even higher derivatives of the scalar field and/or higher curvature terms.

% 4. What we did
With such further extension in mind, we parameterize the effects of gravitational parity violation on the propagation of GWs in a similar way to the PPE framework. 
Our parametrization includes the CNCL model as well as CS gravity. 
We perform a grid survey with the catalog GWTC--1 using our parametrized waveform, as well as the comparison with the analysis using the PPE waveform.
We find that GR is consistent with the data and hence obtain constraints on parity violation in gravity for various PN--order modifications for the first time. 
Our constraint on CS gravity is consistent with the previous work~\cite{Yagi:2017zhb}, while our constraint on the CNCL model is tighter than the previous result by roughly 7 orders of magnitude.

% 5. Organization of the paper.
This paper is organized as follows.
Section~\ref{Sec:waveforms} briefly recapitulates the PPE framework and presents how the gravitational waveform is modified in PV gravity. 
Section~\ref{Sec:Analysis} shows the results of our analysis of LVC open data of the GW catalog using the modified gravitational--wave templates. 
Section~\ref{Sec:Summary} is devoted to summary and discussion. 
We adopt the conventions of~\cite{misner1973gravitation}, in particular for the signature of the metric, Riemann, and Einstein tensors. 
Throughout this paper we use geometric units in which $G = 1 = c$.

\section{Parametrized gravitational waveform}
\label{Sec:waveforms}
\subsection{The PPE waveform}
\label{Sec:PPE}

In this section, we briefly review the PPE waveform following Refs.~\cite{Yunes:2009ke, Tahura:2018zuq}. 
In the PPE framework, the modified waveform is designed by assuming the dominance of the leading PN--order corrections in the binding energy and GW luminosity. 
The frequency domain of the PPE waveform for the inspiral phase of compact binaries is expressed as
\begin{align}
  \label{Eq:PPEtemp}
  \tilde{h} = \tilde{h}^{\rm GR} ( 1 + \alpha_{\rm PPE} u^{a_{\rm PPE}} ) \exp \left( - i \, \beta_{\rm PPE} u^{b_{\rm PPE}} \right) \,, 
\end{align}
where $\alpha_{\rm PPE}$, $\beta_{\rm PPE}$, $a_{\rm PPE}$, and $b_{\rm PPE}$ are the PPE parameters, which represent modifications to the amplitude and the phase, and $\tilde{h}_{\rm GR}$ is the waveform in GR with phase given by
% and its phase is given by
 \begin{align}
   \label{Eq:GRPhase}
   \Psi_{\rm GR} = 2 \pi f t_c - \phi_c - \frac{\pi}{4} + \frac{3}{128} u^{-5} + \cdots
 \end{align}
with frequency $f$, coalescence time $t_c$ and phase $\phi_c$. We have introduced
\[
  u \equiv ( \pi \mathcal{M} f )^{1/3} \,,
\]
where $\mathcal{M} = M \, \nu^{3/5}$ is the chirp mass with the total mass of the binary $M = m_1 + m_2$ and the symmetric mass ratio $\nu = m_1 m_2 / M^2$. 
$a_{\rm PPE}$ and $b_{\rm PPE}$ are the parameters that specify the PN order of the modifications to the amplitude and the phase, respectively. 
In the next subsection, we consider how GWs are modified in PV gravity theories discussed, {\it e.g.} in Refs.~\cite{Crisostomi:2017ugk, Nishizawa:2018srh}, and parametrize the modified templates in a model--independent way.

\subsection{The PV waveform}
\label{Sec:PV}

We decompose the GW $h_{i j}$ as
\[
  h_{i j} = \sum_P h_P e^P_{i j} \,,
\]
where $P$ denotes polarization states and $e^{P}_{i j}$ is the polarization basis. 
For convenience, we choose R and L, which stand for right--handed and left--handed modes, respectively, as the independent bases. They are related to the $+/\times$ mode polarization tensors as
\[
  e^{\rm R}_{i j} = \frac{e^+_{i j} + i \, e^{\times}_{i j}}{\sqrt{2}} \,,
  \qquad
  e^{\rm L}_{i j} = \frac{e^+_{i j} - i \, e^{\times}_{i j}}{\sqrt{2}} \,,
\]
which obey
\[
  \epsilon^{i j k} n_i e^{\rm R, L}_{k l} = i \, \lambda_{\rm R, L} e^{j \, {\rm R, L}}_{\,l} \,,
\]
with $\lambda_{\rm R} = + 1$ and $\lambda_{\rm L} = - 1$; $\epsilon^{i j k}$ is Levi--Civita symbol, {\it e.g.} $\epsilon^{123} = 1$. Similarly, we have
\[
  h_{\rm R} = \frac{h_+ - i \, h_{\times}}{\sqrt{2}} \,,
  \qquad
  h_{\rm L} = \frac{h_+ + i \, h_{\times}}{\sqrt{2}} \,.
\]

Using this decomposition, the linearized equation of motion of GWs on the Friedmann--Lema{\^i}tre--Robertson--Walker background in the CNCL model can be written as~\cite{Nishizawa:2018srh}
\begin{align}
  \left( 1 - \lambda_{\rm P} \, \tilde{k} \, \gamma \right) \, \left( h^{\rm P}_k \right)'' + \left[ 2 - \lambda_{\rm P} \, \tilde{k} \, \left( \gamma + \gamma' \, \mathcal{H}^{-1} \right) \right] \, \mathcal{H} \, \left( h^{\rm P}_k \right)' + \left( 1 - \lambda_{\rm P} \, \tilde{k} \, \delta \right) \, k^2 \, h^{\rm P}_k = 0 \,,
\end{align}
where the prime denotes the differentiation with respect to the conformal time coordinate $\eta$, $k = |\bs{k}|$ is the comoving wavenumber, and $\mathcal{H} = a' / a$ with the scale factor $a = a(\eta)$. 
$\gamma(\eta)$ and $\delta(\eta)$ are time--dependent real--valued parameters of the CNCL model.%
\footnote{In Ref.~\cite{Nishizawa:2018srh}, $\gamma$ and $\delta$ are respectively described as $\alpha$ and $\beta$. However, we use $\gamma$ and $\delta$ in this paper to avoid confusion with the PPE parameters. }
The cut--off energy scale $\Lambda$ is chosen so as for the larger between  $\gamma$ and $\delta$ to be $\mathcal{O}(1)$. 
We introduce the dimensionless wavenumber by
\[
  \tilde{k} \equiv \frac{k}{a \Lambda}.
\]
The CNCL model has been constrained from the arrival--time difference between GWs and photons for GW170817 as~\cite{Nishizawa:2018srh}
\begin{align}
  \label{Eq.constNishizawa}
  \Lambda^{-1} \left|\gamma -\delta \right| \lesssim 10^{-11} \, {\rm km} \,.
\end{align}
Denoting the metric perturbation $h^{\rm P}_k$ by
\[
  h^{\rm P}_k = \hat{A}^{\rm P}_k e^{- i \phi_{\rm P} (\eta)} \,,
\]
with a constant $\hat{A}^{\rm P}_k$, the equation of motion can be rewritten as
\[
  i \phi_{\rm P}'' + ( \phi_{\rm P}' )^2 - \frac{1 - \lambda_{\rm P} \, \tilde{k} \, \delta}{1 - \lambda_{\rm P} \, \tilde{k} \, \gamma} k^2 = - 2 \, i \, \phi_{\rm P}' \frac{\chi'}{\chi} \,,
\]
where we define
\[
  \chi \equiv a \sqrt{1 - \lambda_{\rm P} \, \tilde{k} \, \gamma} \,.
\]
Since the frequency of GWs is high compared with the cosmological time scale, {\it i.e.} $\mathcal{H} \ll k$, we assume slow time variation
\[
  \phi_{\rm P}'' \ll ( \phi_{\rm P}' )^2 \,,
\]
with
\begin{align}
  \label{Eq:WeakCS}
  \tilde{k} \ll 1 \,,
\end{align}
which corresponds to the weak PV coupling.
Under the above assumptions, one finds
\[
  \phi_{\rm P} = \phi^{\rm P}_{\rm Re} + i \, \phi^{\rm P}_{\rm Im} \,,
\]
where
\begin{align}
\label{Eq:phiIP}
  \phi^{\rm P}_{\rm Im} &= \ln \left[ \frac{\chi (\eta_{\rm s} )}{\chi (\eta_0 )} \right] \simeq \ln \frac{a_{\rm s}}{a_0} + \lambda_{\rm P} \frac{\pi a_0 f}{\Lambda} \left( \frac{\gamma_0}{a_0} - \frac{\gamma_{\rm s}}{a_{\rm s}} \right) \,, \\
  \phi^{\rm P}_{\rm Re} &= \pm k \int_{\eta_{\rm s}}^{\eta_0} \dd \eta \sqrt{\frac{1 - \lambda_{\rm P} \, \tilde{k} \, \delta}{1 - \lambda_{\rm P} \, \tilde{k} \, \gamma}} \simeq \pm \left[ 2 \pi a_0 f ( \eta - \eta_{\rm s} ) + 2 \lambda_{\rm P} \frac{\pi^2 a_0^2 f^2}{\Lambda} \int_{\eta_{\rm s}}^{\eta_0} \frac{\dd \eta}{a} \left( \gamma - \delta \right) \right] \,,
\end{align}
with $f = k / ( 2 \pi a_0 )$. The subscripts ``$0$'' and ``$s$'' refer to the quantities evaluated at the present and $\eta = \eta_{\rm s}$, respectively.
Note that the first term in the right--hand side of Eq.~\eqref{Eq:phiIP} can be absorbed by the shift of the source distance since this is parity symmetric and constant, and thus we neglect it in the following discussion. 
Therefore, the circular polarization modes of GWs are modified as
\begin{align}
  \tilde{h}^{\rm CNCL}_{\rm R, L} = \tilde{h}^{\rm GR}_{\rm R, L} \left( 1 + \lambda_{\rm R, L} \dA^{\rm CNCL} \right) \exp \left( - i \lambda_{\rm R, L} \dPhi^{\rm CNCL} \right) \,,
\end{align}
with
\begin{align}
  \dA^{\rm CNCL} = \frac{\pi f}{\Lambda} \left( \gamma_0 - \frac{\gamma_{\rm s}}{a_{\rm s}} \right) \,,
  \quad
  \dPhi^{\rm CNCL} = \frac{2 \pi^2 f^2}{\Lambda} \int_{\eta_{\rm s}}^{\eta_0} \frac{\dd \eta}{a} \left( \gamma - \delta \right) \,,
\end{align}
where we set $a_0 = 1$. 
The opposite sign depending on the chirality associated with $\dA^{\rm CNCL}$($\dPhi^{\rm CNCL}$) causes the so--called amplitude(phase) birefringence. 
In terms of the PN order, $\dA^{\rm CNCL}$ and $\dPhi^{\rm CNCL}$ are, respectively, 1.5PN-- and 5.5PN--order corrections. 
We should notice that the weak PV coupling approximation [Eq.~\eqref{Eq:WeakCS}] does not always mean the corrections $\dA^{\rm CNCL}$ and $\dPhi^{\rm CNCL}$ are much smaller than unity because the former depends on the difference in $\gamma$ between the source and us and the latter is evaluated by integrating the parameters $\gamma$ and $\delta$ over the propagating distance.

With the same philosophy as the PPE framework, the gravitational waveforms in PV gravity can be parametrized as
\begin{align}
  \label{Eq:tempPV}
  \tilde{h}_{\rm R, L} = \tilde{h}^{\rm GR}_{\rm R, L} ( 1 + \lambda_{\rm R, L} \dA ) e^{- i \lambda_{\rm R, L} \dPhi} \,,
\end{align}
with
\begin{align}
  \label{Eq:deltaAPhi}
  \dA = \alpha_{\rm PV} u^{a_{\rm PV}} \,,
  \quad
  \dPhi = \beta_{\rm PV} u^{b_{\rm PV}} \,,
\end{align}
where $\alpha_{\rm PV}$, $\beta_{\rm PV}$, $a_{\rm PV}$, and $b_{\rm PV}$ are the parameters introduced to describe the amplitude and the phase birefringences for an arbitrary PN order.%
\footnote{Obviously, the CNCL model corresponds to
\[
  \alpha_{\rm PV}^{\rm CNCL} = \frac{1}{\mathcal{M} \Lambda} \left( \gamma_0 - \frac{\gamma_{\rm s}}{a_{\rm s}} \right),
  \quad
  a_{\rm PV}^{\rm CNCL} = 3 \,,
  \quad
  \beta_{\rm PV}^{\rm CNCL} = \frac{2}{\mathcal{M}^2 \Lambda} \int_{\eta_{\rm s}}^{\eta_0} \dd \eta \left( \frac{\gamma}{a} - \frac{\delta}{a} \right),
  \quad
  b_{\rm PV}^{\rm CNCL} = 6.
\]}
The PN order of the correction to the amplitude is $a_{\rm PV}/2$, while that to the phase is $(b_{\rm PV} + 5)/2$. We emphasize again $\dA$ and $\dPhi$ can be larger than unity because of a long propagating distance. 

\begin{figure}[tbp]
  \begin{center}
    \includegraphics[width=140mm]{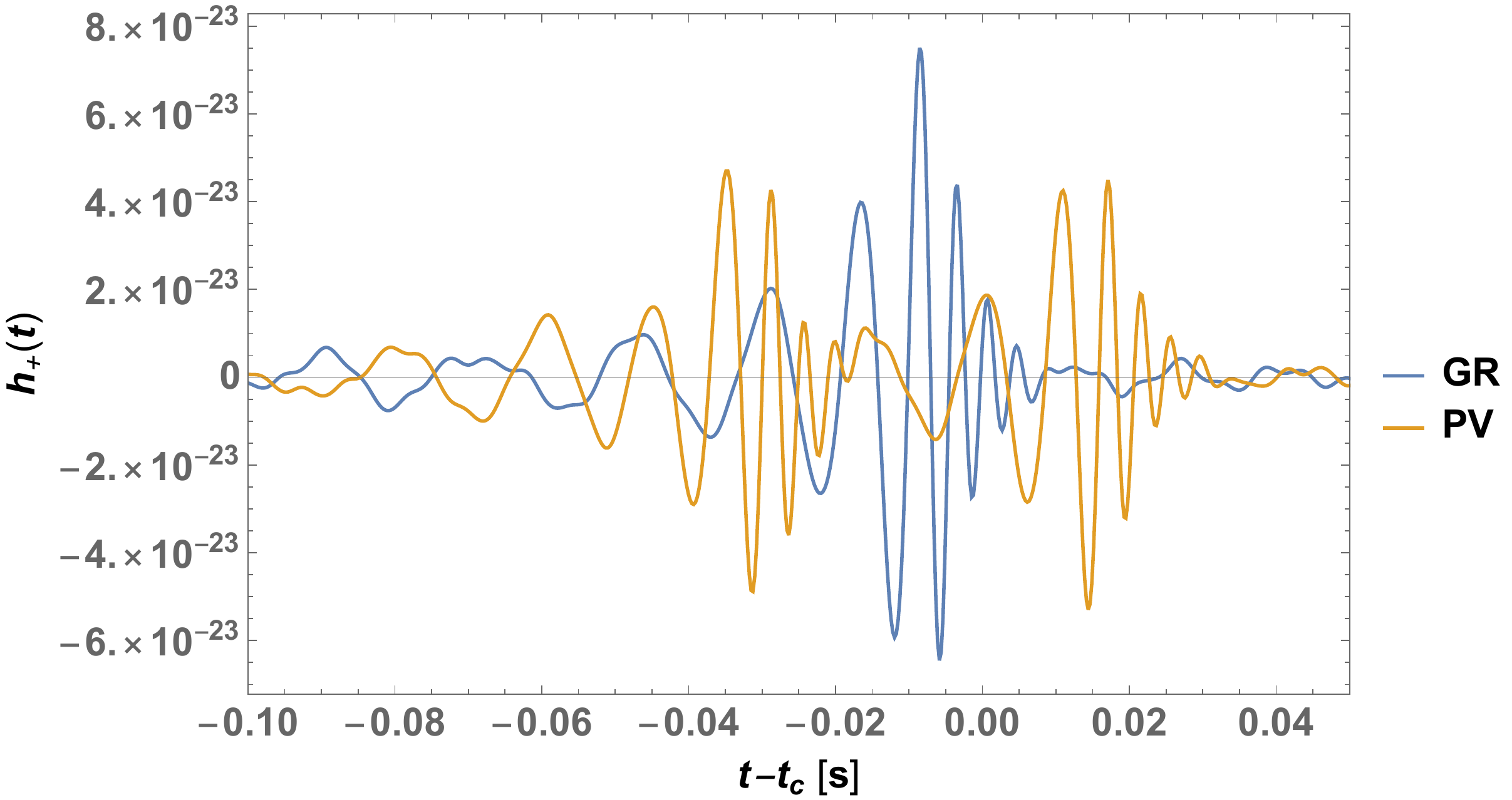}
  \end{center}
  \caption{An example of the plus--mode whitened templates with $t_c$ fixed. The blue--solid line is the GR template for a GW150914--like event, while the orange--dashed line shows the modified template Eq.~\eqref{Eq:modhp} with the parameters as $(\iota, \alpha_{\rm PV}, \beta_{\rm PV}, b_{\rm PV}) = ( \pi/2, 0, 3 \times 10^2, 3)$. There are two merger--ringdown signals in the modified template because the plus mode consists of the superposition of a pair of circular polarization modes, which consists of superluminal and subluminal modes.}
  \label{Fig:ExampleTemp}
\end{figure}

The $+/\times$ mode polarizations become
\begin{align}
  \label{Eq:modhp}
  \tilde{h}_+
  &= \frac{\tilde{h}_{\rm R} + \tilde{h}_{\rm L}}{\sqrt{2}} 
  = \left( \tilde{h}^{\rm GR}_+  - i \, \tilde{h}^{\rm GR}_{\times} \dA \right) \cos \dPhi - \left( \tilde{h}^{\rm GR}_{\times} + i \, \tilde{h}^{\rm GR}_+ \dA \right) \sin \dPhi \,, \\
  \label{Eq:modhc}
  \tilde{h}_{\times}
  &= i \frac{\tilde{h}_{\rm R} - \tilde{h}_{\rm L}}{\sqrt{2}} 
  = \left( \tilde{h}^{\rm GR}_{\times} + i \, \tilde{h}^{\rm GR}_+ \dA \right) \cos \dPhi + \left( \tilde{h}^{\rm GR}_+ - i \, \tilde{h}^{\rm GR}_{\times} \dA \right) \sin \dPhi \,.
\end{align}
In general, these two modes are not orthogonal. 
Therefore, for the analysis in the next section, we orthonormalize the template by using Eq.~\eqref{Eq:orthogonalh}.
Figure~\ref{Fig:ExampleTemp} shows an extreme example of the modified template with the inclination angle of the source $\iota = \pi/2$, $\alpha_{\rm PV} = 0$, $\beta_{\rm PV} = 3 \times 10^2$, and $b_{\rm PV} = 3$ for a GW150914--like signal.%
\footnote{The 4PN--order phase birefringence results in a superposition of two merger--ringdown waveforms, which is similar to strongly lensed GWs. However, in the PV case the respective circular polarization modes carry only one of the two, while both modes become a superposition for lensed GWs. Therefore, such a modification in PV gravity is distinguishable from the lensed GWs.}

In CS gravity, the modification reduces to~\cite{Yagi:2017zhb}
\begin{align}
  \tilde{h}_+^{\rm CS} = \tilde{h}^{\rm GR}_+  - i \, \tilde{h}^{\rm GR}_{\times} \dA^{\rm CS} \,,
  \quad
  \tilde{h}_{\times}^{\rm CS} = \tilde{h}^{\rm GR}_{\times} + i \, \tilde{h}^{\rm GR}_+ \dA^{\rm CS} \,,
\end{align}
where
\begin{align}
  \dA^{\rm CS} = \pi f \left( \dot \vartheta_0 - ( 1 + z ) \dot \vartheta_{\rm s} \right) \simeq \pi f \left( \dot \vartheta_0 - \frac{\ddot \vartheta_0}{H_0} \right) z \,
\end{align}
with the Hubble constant $H_0$, the redshift of the source $z$, and the scalar field $\vartheta = \vartheta(t)$. The current constraint on $|\dot \vartheta_0|$ obtained from the binary pulsar observations by neglecting $\ddot \vartheta_0$ is~\cite{Yunes:2008ua,AliHaimoud:2011bk} 
\begin{align}
  \label{Eq:currentCS}
  |\dot \vartheta_0| \lesssim 0.4 \, {\rm km} \,.  
\end{align}

\section{Analysis}
\label{Sec:Analysis}

\subsection{Setup}
\label{Sec:Setup}

We employ the waveform Eq.~\eqref{Eq:tempPV} with Eq.~\eqref{Eq:deltaAPhi} as templates to be matched with the strain data of Hanford and Livingston taken from the confident detections cataloged in GWTC--1~\cite{Abbott:2019ebz}. 
Using the KAGRA Algorithmic Library (KAGALI)~\cite{Oohara:2017aix}, we evaluate the likelihood, following the standard procedure of the matched filtering~\cite{maggiore2008gravitational, 2009agwd.book.....J, Creighton:2011zz}. 
We adopt the published noise power spectrum for each event~\cite{Abbott:2019ebz}. 
The minimum and the maximum frequencies, $f_{\rm min}$ and $f_{\rm max}$, of the datasets used in the analysis are summarized in Table~\ref{Table:bstfitparameters}. 
As the GR waveform $\tilde{h}_{\rm GR}$, we adopt IMRPhenomD~\cite{Husa:2015iqa, Khan:2015jqa}, which is an up--to--date version of inspiral--merger--ringdown (IMR) phenomenological waveform for binary BHs with aligned spins.

In this paper, since we are not interested in estimating the sky position of the source, we choose the amplitudes' ratio of the $+/\times$--mode of the respective detectors, {\it i.e.} Eq.~\eqref{Eq:orthogonalh}, so that the sky position of the waveform template is set to the best--fitting values in an analytical way. 
The inclination of the source orbit, on the other hand, is an important parameter in the PV modification, since it is (approximately) degenerated with the magnitudes of birefringences, $\alpha_{\rm PV}$ and/or $\beta_{\rm PV}$. 
Moreover, if the source is a face--on binary, only one of the circular polarization modes can propagate to us. 
In this case, the PV template, Eq.~\eqref{Eq:tempPV}, reduces to the PPE template, Eq.~\eqref{Eq:PPEtemp}, and thus PV modifications can never be tested. 
To test the PV modifications, GW events from nearly edge--on binaries, for which both circular polarization modes can propagates with almost the same amplitudes, are required. 

First, we implement a grid survey to find the ``best--fitting parameters'' of GR templates for each event varying the parameters $\mathcal{M}$, $\nu$, $\chi_1$, and $\chi_2$ as well as $t_c$ and $\phi_c$. 
In GR, we can analytically set the inclination as the best--fitting value. 
The results in the detector frame are summarized in Table~\ref{Table:bstfitparameters}. 
Next, we calculate the likelihood for the modified templates around the GR best--fitting parameters. 
Since the inclination of the source is important to test the PV modifications, we vary the inclination angle as well as the additional parameters $\alpha_{\rm PV}$, $\beta_{\rm PV}$, $a_{\rm PV}$ and $b_{\rm PV}$. 
A result for GW150914 suggests that even if both birefringences are taken into consideration at the same time, the constraints on the parameters do not change significantly (see Appendix~\ref{Sec:bothalphabeta}). 
Therefore, we focus only on either the amplitude or the phase birefringences in order to further reduce the number of parameters. 
The former includes the CS modification, while the latter includes the CNCL model with negligible amplification. 
For both birefringences, in general, lighter chirp mass events can impose stronger constraints for negative PN corrections. 
This is because events with a lighter chirp mass have a longer inspiral phase and the negative PN--order corrections become more efficient in the early inspiral phase.
On the other hand, to constrain the positive PN--order corrections, {\it e.g.} 1.5PN and 5.5PN birefringences, heavier chirp mass events are preferred, because the modifications become more significant after the late--inspiral phase. 
Moreover, as is well known, there is an approximate degeneracy among the mass ratio and the spins in the inspiral phase waveform. 
In fact, we find it unnecessary to take into account both variations in our calculations. 
Therefore, here we fix the spins to save computational costs.

\begin{table}[tb]
  \begin{center}
    \caption{The GR ``best--fitting parameters'' in the detector frame, adopted from Ref.~\cite{Yamada:2019zrb}.}
  \begin{tabular}{c c c c c c c} \hline
    Event & $(f_{\rm min}, f_{\rm max})$/Hz & $\mathcal{M}/M_{\odot}$ & $\nu$ & $\chi_1$ & $\chi_2$ & SNR \\
    \hline
    \hline
    GW150914 & (20, 1024) & 31.2 & 0.249 &  0.79 & -1.00 & 24.4 \\
    GW151012 & (20, 1024) & 18.0 & 0.249 & -0.36 &  0.22 & 9.00 \\
    GW151226 & (20, 1024) & 9.70 & 0.211 &  0.50 & -0.46 & 12.0 \\
    GW170104 & (20, 1024) & 24.9 & 0.243 & -0.87 &  1.00 & 13.2 \\
    GW170608 & (20, 1024) & 8.48 & 0.250 &  0.47 & -0.49 & 15.4 \\
    GW170729 & (20, 1024) & 48.4 & 0.201 &  0.74 & -1.00 & 10.4 \\
    GW170809 & (20, 1024) & 30.5 & 0.228 &  0.72 & -1.00 & 12.0 \\
    GW170814 & (20, 1024) & 26.9 & 0.246 &  0.85 & -1.00 & 16.2 \\
    GW170817 & (23, 2048) & 1.20 & 0.231 & -0.41 &  0.96 & 32.2 \\
    GW170818 & (16, 1024) & 32.9 & 0.250 &  0.98 & -1.00 & 10.5 \\
    GW170823 & (20, 1024) & 39.3 & 0.249 &  0.99 & -1.00 & 11.5 \\
    \hline
  \end{tabular}
    \label{Table:bstfitparameters}
  \end{center}
\end{table}

\subsection{Results}

\begin{figure}[tbp]
  \begin{center}
    \includegraphics[width=140mm]{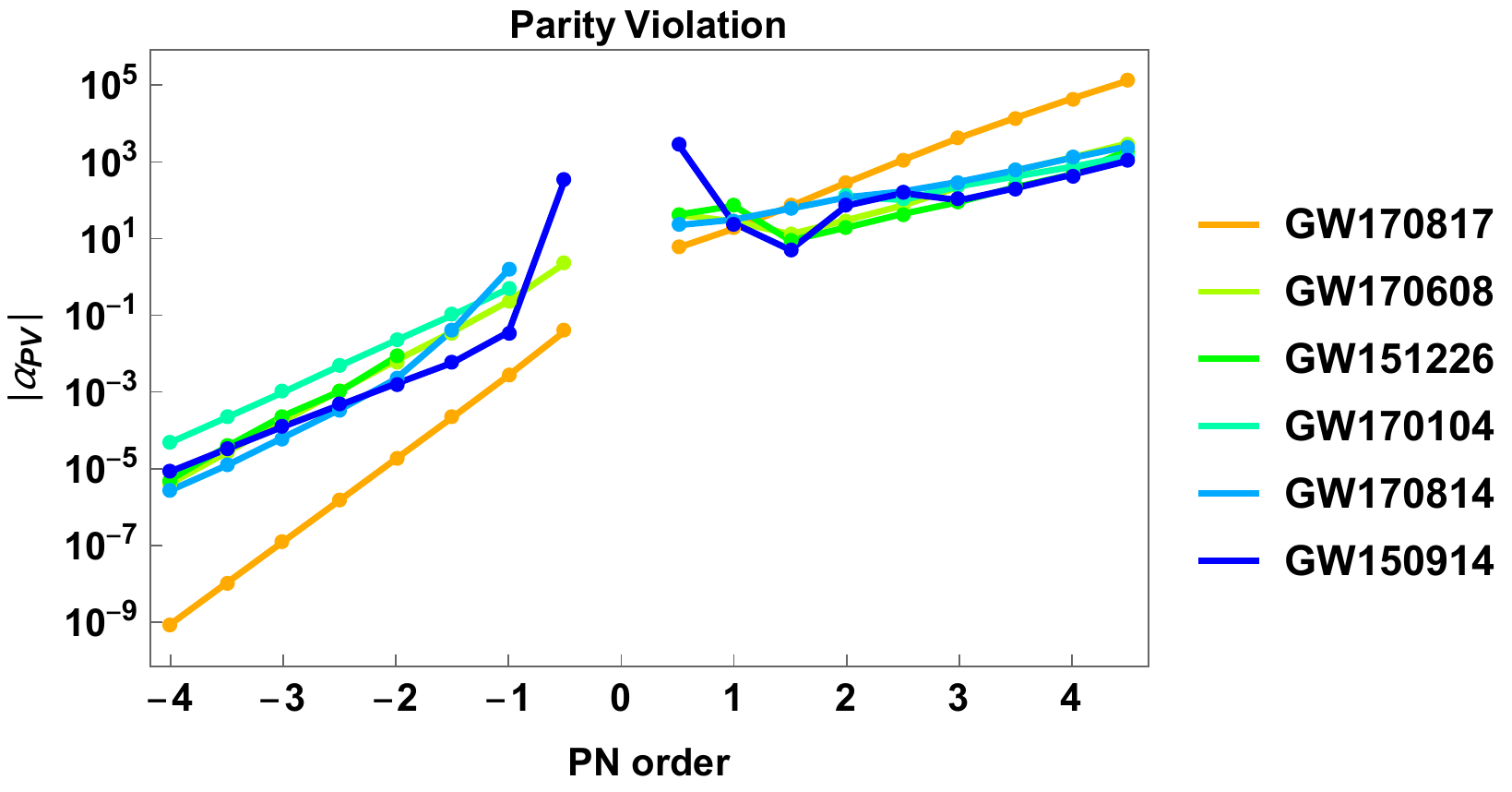}
    \includegraphics[width=140mm]{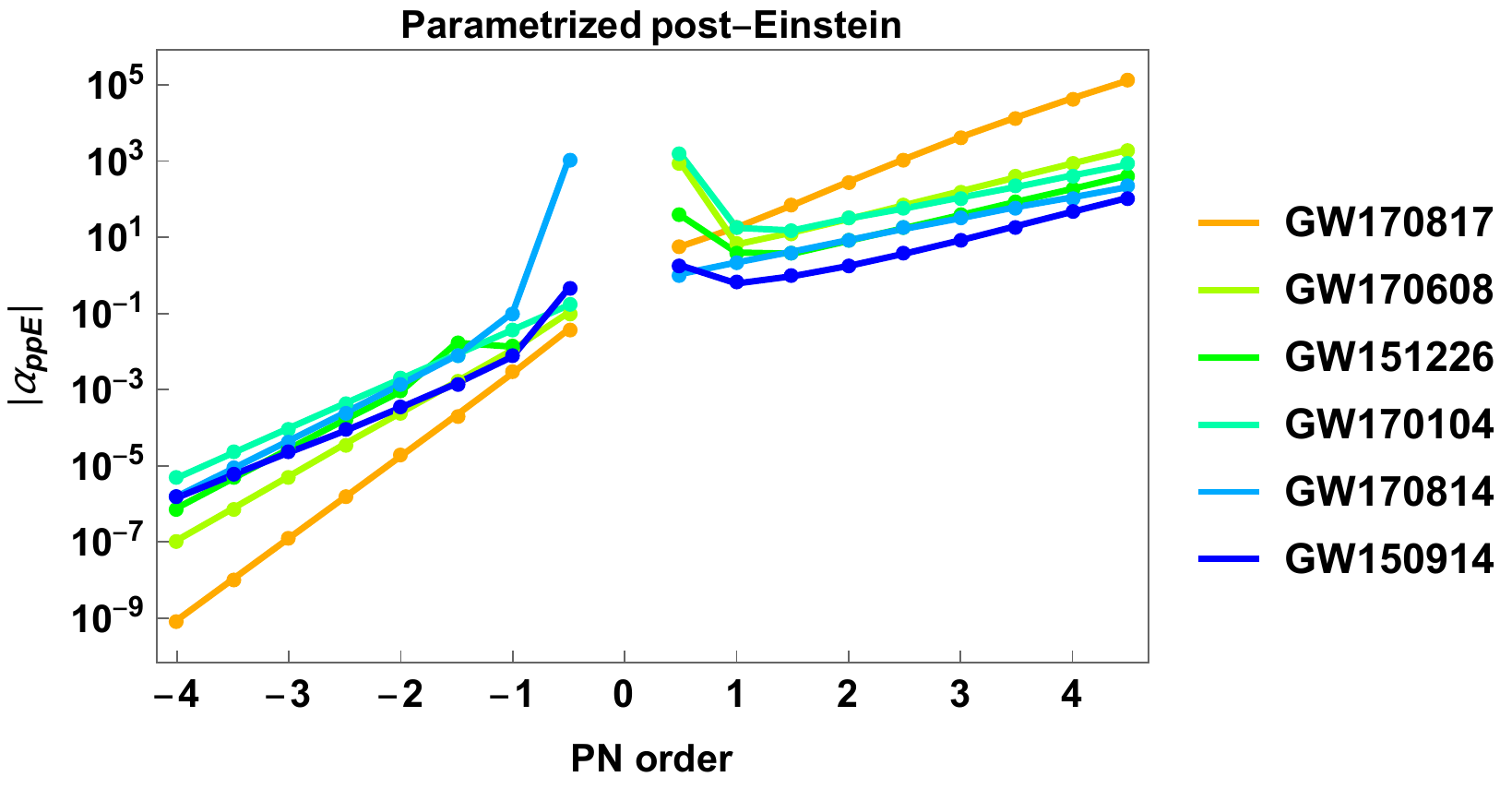}
  \end{center}
  \caption{Upper panel: the 90\% CL constraints on $\alpha_{\rm PV}$. Lower panel: the same on $\alpha_{\rm PPE}$.}
  \label{Fig:alphaPVvsPPE}
\end{figure}

\begin{figure}[tbp]
  \begin{center}
    \includegraphics[width=140mm]{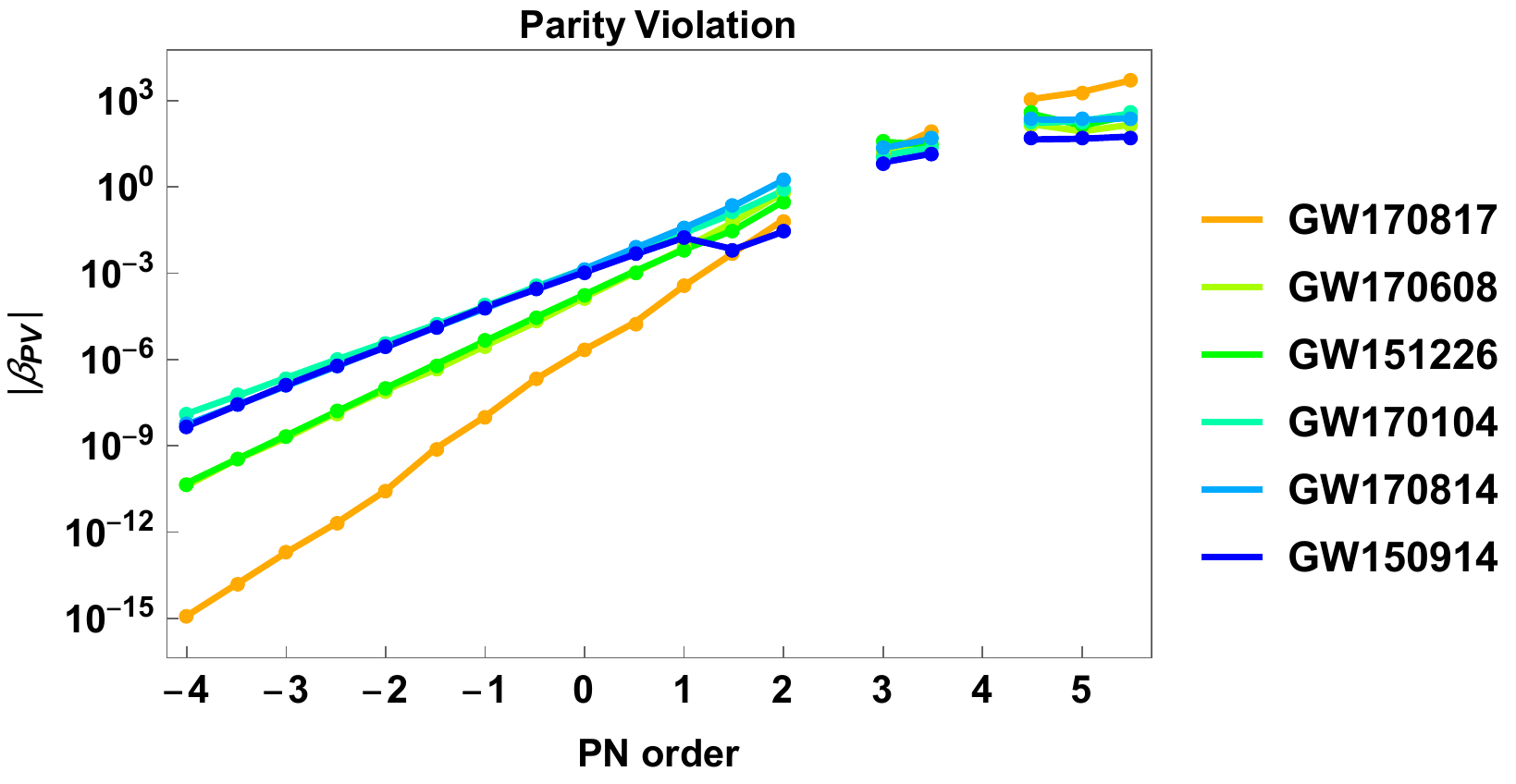}
    \includegraphics[width=140mm]{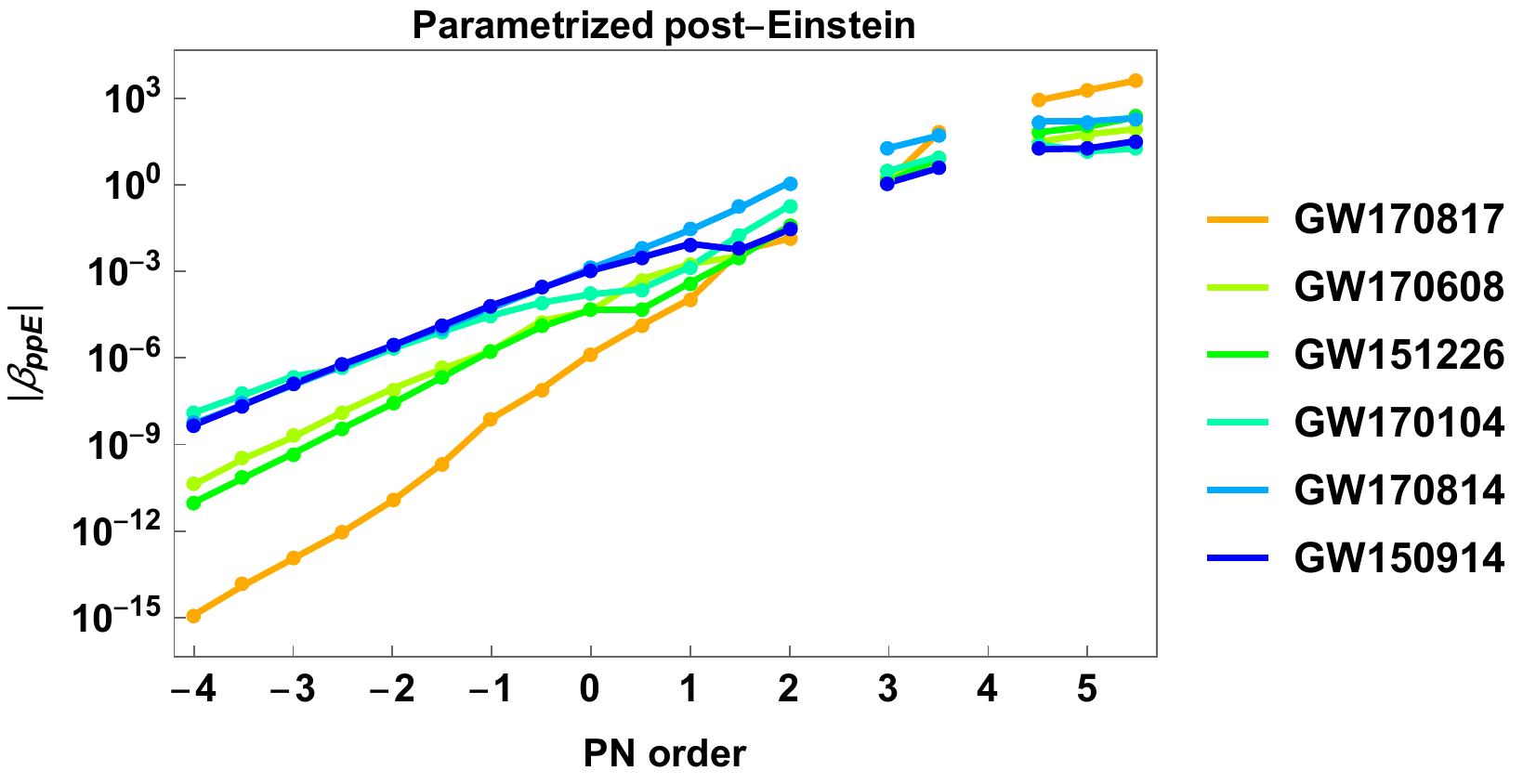}
  \end{center}
  \caption{Upper panel: the 90\% CL constraints on $\beta_{\rm PV}$. Lower panel: the same on $\beta_{\rm PPE}$.}
  \label{Fig:betaPVvsPPE}
\end{figure}

For the amplitude birefringence, it is difficult to constrain the parameter $\alpha_{\rm PV}$ for the 0PN order from only GW observations, since there is a degeneracy among the source distance, the inclination, and $\alpha_{\rm PV}$. 
Similarly, near the 0PN order, such as -0.5PN and 0.5PN orders, these parameters also weakly degenerate. Therefore, the constraints at such PN orders tend to be weaker. 
If we find the host galaxy of the source like GW170817, $\alpha_{\rm PV}$ at the 0PN order can be constrained as $\alpha_{\rm PV} < \mathcal{O}(1)$. This is because if $\alpha_{\rm PV}$ at the 0PN order is much larger than unity and $\beta_{\rm PV} = 0$, then the $+/\times$--modes are flipping and much amplified as can be seen from Eqs.~\eqref{Eq:modhp} and \eqref{Eq:modhc}. Such large amplification causes underestimation of the distance that contradicts the multi--messenger observation results. On the other hand, if $\alpha_{\rm PV} < \mathcal{O}(1)$, then $\alpha_{\rm PV}$ completely degenerates with the inclination angle, and hence it cannot be constrained only from the GW observations.

For the phase birefringence, there are degeneracies at the 2.5PN and 4PN orders. 
At the 2.5PN order, $\beta_{\rm PV}$ degenerates with the coalescence phase $\phi_c$ and in principle this degeneracy cannot be solved (see Appendix~\ref{Sec:constdphi}). 
At the 4PN order, on the other hand, the situation is different. 
First, if the GW source is completely face--on, {\it i.e.} $\iota = 0$, so that one of two circular polarization modes vanishes, then the PV template [Eq.~\eqref{Eq:tempPV}] reduces to the PPE template [Eq.~\eqref{Eq:PPEtemp}]. 
In this case, the parameter $\beta_{\rm PV}$ completely degenerates with the coalescence time $t_c$ and thus we cannot constrain the modification. 
In principle, this degeneracy is gradually resolved as the inclination increases. 
Even for nearly face--on binaries with non--negligible amplitude of the weaker circular polarization mode, it is still difficult to constrain the parameter only from the GW signals because of the difficulty in distinguishing the {\it secondary wave} from the noise. 
Please see Appendix~\ref{Sec:4PNnoise} for the detailed discussion. 
Therefore, GW signals from nearly edge--on binaries are required to test the 4PN--order phase birefringence in PV gravity.

To obtain the constraints, we calculate the likelihood of the best--fitting template for each PN--order modification. 
The likelihood can be regarded as the unnormalized posterior distribution when the prior of $\alpha_{\rm PV}$($\beta_{\rm PV}$) is uniform as well as those of $\mathcal{M}$ and $\nu$.
Here, we use the likelihood maximized for the other GR parameters instead of the marginalized likelihood. This is because the marginalized likelihood depends on the prior distributions of parameters and the error caused by this naive treatment would be within the arbitrariness in the choice of the prior distribution.
Thus, we integrate it with respect to $\alpha_{\rm PV}$($\beta_{\rm PV}$) for each PN order to evaluate 90\% confidence level (CL) constraints. 
As expected, the results shown below indicate that the constraints tend to be stronger for lighter chirp mass events for the negative PN--order modifications, while the heavier mass events are stronger for the positive PN orders.. 
Moreover, we find that the constraints on PV gravity tend to be weaker than those on the PPE modifications by a factor. 
This is because the inclination in the PPE waveform is completely degenerate with $\phi_c$, while it is not in the PV modifications. 
Therefore, in the PV case the effective number of parameters is larger than in the PPE framework, and hence the likelihood tends to be larger.

First, we consider the amplitude birefringence by varying the PN order from -4 to 4.5. The 1.5PN--order corresponds to CS gravity. 
The upper panel of Fig.~\ref{Fig:alphaPVvsPPE} shows the 90\% CL constraints on $\alpha_{\rm PV}$ for each event. The lower panel shows the 90\% CL constraints on $\alpha_{\rm PPE}$ for comparison.%
\footnote{Since we consider the modifications of GWs during propagation, in this analysis we never truncate the PPE modifications in Eq.~\eqref{Eq:PPEtemp} even in merger--ringdown phase. Therefore, the constraints on the PPE parameters we obtained are slightly different from the previous works, such as Refs.~\cite{LIGOScientific:2019fpa,Yunes:2016jcc}.}
Here, we show only six events possessing relatively high SNR.
We obtain an upper bound on CS gravity from the constraints at the 1.5PN order as
\begin{align}
  \ell_{\rm CS} \equiv \left| \dot \vartheta_0 - \frac{\ddot \vartheta_0}{H_0} \right| \lesssim 10^3 \left( \frac{z}{z_{\rm LVC}} \right)^{-1} \, {\rm km} \,,
\end{align}
where $z_{\rm LVC}$ is the mean value of the redshift that LVC estimated in Ref.~\cite{LIGOScientific:2018mvr}.
If we neglect $\ddot \vartheta_0$, this constraint becomes roughly $\left|\dot \vartheta_0\right| \lesssim 10^3 \, {\rm km} $, which is still much weaker than the constraint from the binary pulsar observations [Eq.~\eqref{Eq:currentCS}] as discussed in Ref.~\cite{Yagi:2017zhb} with the Fisher analysis.

Next, the phase birefringence is considered, varying the PN order from -4 to 5.5. 
The upper panel of Fig.~\ref{Fig:betaPVvsPPE} shows the 90\% CL constraints on $\beta_{\rm PV}$ for each event. The lower panel shows the 90\% CL constraints on $\beta_{\rm PPE}$ for comparison. Similarly to Fig.~\ref{Fig:alphaPVvsPPE}, we show only six events possessing relatively high SNR.
We obtain upper bounds on the CNCL model as
\begin{align}
  \ell_{\rm CNCL} \equiv \frac{1}{d_{\rm LVC} \Lambda} \left| \int_{\eta_{\rm s}}^{\eta_0} \dd \eta \left( \frac{\gamma}{a} - \frac{\delta}{a} \right) \right| \lesssim 10^{-18}  \, {\rm km} \,,
\end{align}
where $d_{\rm LVC}$ is the mean value source distance of LVC estimation in Ref.~\cite{LIGOScientific:2018mvr}. Neglecting the time dependence of $\gamma$ and $\delta$, this can be rewritten as
\begin{align}
  \label{Eq.constCNCL}
  \Lambda^{-1} \left|\gamma - \delta \right| \lesssim 10^{-18} \left( \frac{t_0 - t_{\rm s}}{d_{\rm LVC}} \right)^{-1} \, {\rm km} \,,
\end{align}
which improves the existing bound [Eq.~\eqref{Eq.constNishizawa}] by roughly 7 digits.

$\ell_{\rm CS}$ and $\ell_{\rm CNCL}$ that make $\dA$ and $\dPhi$ to be $\mathcal{O}(1)$ are estimated as
\begin{align}
\ell_{\rm CS} &\simeq \frac{1}{2 \pi f z} \approx 2 \times 10^3 \left( \frac{f}{150 \, {\rm Hz}} \right)^{-1} \left( \frac{z}{0.1} \right)^{-1} \, {\rm km} \,, \\
\ell_{\rm CNCL} &\simeq \frac{1}{4 \pi^2 f^2 d} \approx 3 \times 10^{-18} \left( \frac{f}{150 \, {\rm Hz}} \right)^{-2} \left( \frac{d}{1 \, {\rm Gpc}} \right)^{-1} \, {\rm km} \,,
\end{align}
where $d$ is the source distance. These estimates confirm that the constraints on $\ell_{\rm CS}$ and $\ell_{\rm CNCL}$ obtained above are reasonable.
The 90\% CL constraints from each event are summarized in Table~\ref{Table:constraints}. In that table, ``--'' means that 
the likelihood never decreases enough because of the noise and the uncertainty of the amplitude estimate.

\begin{table}[tb]
  \begin{center}
    \caption{90\% CL constraints on the CS coupling and the CNCL model. `--' means that the likelihood never decreases enough because of the noise and the uncertainty of the amplitude estimate.} 
    \begin{tabular}{c c c c c} \hline
    Event & $\alpha^{\rm CS}$ & $\ell_{\rm CS} / {\rm km}$ & $\beta^{\rm CNCL}$ & $\ell_{\rm CNCL} / (10^{-18} \, {\rm km})$ \\
    \hline
    \hline
    GW150914 & 5.05964 & 2585.88 & 56.9279 & 4.43563 \\
    GW151012 & 7695.15 & 974597 & 153.954 & 1.63397 \\
    GW151226 & 8.93474 & 1422.34 & 312.751 & 2.31172 \\
    GW170104 & {\text{--}} & {\text{--}} & 365.191 & 8.08242 \\
    GW170608 & 13.0157 & 2327.93 & 148.741 & 1.1806 \\
    GW170729 & {\text{--}} & {\text{--}} & 522.087 & 15.2002 \\
    GW170809 & {\text{--}} & {\text{--}} & 390.002 & 12.4564 \\
    GW170814 & 60.5647 & 20081.5 & 247.156 & 10.5671 \\
    GW170817 & 70.0234 & 12384.3 & 5185.29 & 6.57038 \\
    GW170818 & {\text{--}} & {\text{--}} & 148.441 & 5.41785 \\
    GW170823 & {\text{--}} & {\text{--}} & 187.229 & 5.25869 \\
    \hline
  \end{tabular}
    \label{Table:constraints}
  \end{center}
\end{table}

\section{Summary and discussion}
\label{Sec:Summary}

In this paper, we discussed the influence of parity violation of gravity to GWs during propagation in a model--independent way by parametrizing the modification of gravitational waveforms. 
Our parametrization includes the CNCL model as well as CS gravity.
The effect of the gravitational parity violation on the gravitational waveform is large when the GW source is nearly edge--on, while our parametrized waveform reduces to the parity--symmetric one, {\it i.e.} the PPE waveform, if the source is nearly face--on.

Furthermore, under the parametrization we perform a grid survey to test such modification by using the LIGO/Virgo O1/O2 catalog. 
We find that general relativity is consistently preferred for those events and obtain constraints on the parity violation of gravity. 
The constraint on CS gravity is $|\dot \vartheta_0| \lesssim 10^3 \, {\rm km}$, which is still weaker than the current constraint obtained from binary pulsar observations. 
On the other hand, the constraint on the CNCL model is improved by roughly 7 digits from the current upper bound.

It turned out to be difficult to constrain the 0PN--order amplitude birefringence and the 2.5PN-- and 4PN--order phase birefringences from GWTC--1. 
The 2.5PN phase birefringence completely degenerates with the coalescence phase $\phi_c$ and thus this cannot be constrained only from the GW observations. 
On the other hand, one of the main reason for the difficulties in constraining the 0PN amplitude and the 4PN phase birefringences is the degeneracy in the inclination angle.
To solve such a degeneracy, it may be helpful to take account of higher multipole modes in the GR waveform, which may break the inclination--distance degeneracy. 
The relative importance of the higher multipole modes may increase as the system asymmetry increases such as unequal masses, unequal spin magnitudes, and the precession due to misaligned spins (see for example Ref.~\cite{Arun:2008kb}). 
Therefore, a large mass ratio event, such as GW190412, may play an interesting role to solve the degeneracy~\cite{LIGOScientific:2020stg}. 
This is left as future work.

\section*{Acknowledgments}

We would like to thank the ``29th Workshop on General Relativity and Gravitation in Japan (JGRG29)'' during which part of this work was developed.
This work was supported by JSPS KAKENHI Grant Number JP17H06358 (and also JP17H06357), A01: {\it Testing gravity theories using gravitational waves}, as a part of the innovative research area, ``Gravitational wave physics and astronomy: Genesis''. 
T.T. acknowledges support from JSPS KAKENHI Grant No. JP20K03928.
We are also grateful to the members of the A01 group for useful discussions and comments.

\appendix

\section{Degeneracy between the 2.5PN phase birefringence and the coalescence phase}
\label{Sec:constdphi}
  
Let us define
\begin{align}
  \tilde{\mathcal{A}} \equiv \left( \tilde{h}^{\rm GR}_+  - i \, \tilde{h}^{\rm GR}_{\times} \dA \right) \,,
  \quad
  \tilde{\mathcal{B}} \equiv \left( \tilde{h}^{\rm GR}_{\times} + i \, \tilde{h}^{\rm GR}_+ \dA \right) \,.
\end{align}
Thus, the modified GWs [Eqs.~\eqref{Eq:modhp}--\eqref{Eq:modhc}] are
\begin{align}
\label{Eq:hmodplus}
  \tilde{h}_+ &= \tilde{\mathcal{A}} \cos \dPhi + \tilde{\mathcal{B}} \sin \dPhi \,, \\
\label{Eq:hmodcross}
  \tilde{h}_{\times} &= \tilde{\mathcal{B}} \cos \dPhi - \tilde{\mathcal{A}} \sin \dPhi \,.
\end{align}
We define the inner--product of two complex quantities $\tilde P(f)$ and $\tilde Q(f)$ as
\begin{align}
  \label{Eq:innerproduct}
  (P, Q) = 4 {\rm Re} \int_0^{\infty} df \frac{P \, Q^*}{S_n} \,,
\end{align}
$S_n$ is the noise power spectrum density.
One can show that $(\mathcal{A}, \mathcal{B}) = 0$. Note that the two modes of the modified GWs are not orthogonal in general. One can find the orthonormalized waveform as
\begin{align}
  \label{Eq:orthogonalh}
  \tilde{h}_{\rm \pm}^{\rm O} = \sum_{{\rm P} = +, \times} \frac{1}{\sqrt{c_{\rm \pm}}} \bs{e}_{\rm \pm P} \tilde{h}_{\rm P} \,,
\end{align}
where $c_{\rm \pm}$ and $\bs{e}_{\rm \pm P}$ are eigenvalues and unit eigenvectors of $\tilde{h}_{\rm P}$. 
Defining 
\begin{align}
  C_+^2 \equiv ( h_+, h_+ ) \,,
  \quad
  C_{\times}^2 \equiv ( h_{\times}, h_{\times} ) \,,
  \quad
  C_{\rm m}^2 \equiv ( h_+, h_{\times} ) \,,
\end{align}
one can express $c_{\rm \pm}$ and $\bs{e}_{\rm \pm P}$ as
\begin{align}
  c_{\rm \pm} &= \frac12 \left( C_+^2 + C_{\times}^2 \pm \sqrt{( C_+^2 - C_{\times}^2 )^2 + 4 C_{\rm m}^2} \right) \,, \\  \label{Eq:eigenvec}
\bs{e}_{\rm \pm P} &= \frac{\bs{v}_{\rm \pm P}}{|\bs{v}_{\rm \pm P}|} \,,
  \quad
\bs{v}_{\rm \pm P} \equiv \left( 2 C_{\rm m}^2, - C_+^2 + C_{\times}^2 \mp \sqrt{( C_+^2 - C_{\times}^2 )^2 + 4 C_{\rm m}^2} \right) \,.
\end{align}

In order to find the degeneracy between the 2.5PN phase birefringence and the phase shift $\phi_c$, we focus on the case of $b = 0$, {\it i.e.} $\dPhi$ is constant. 
Immediately, from Eqs.\eqref{Eq:hmodplus} and \eqref{Eq:hmodcross} one can see that this is nothing but a constant rotation of the coordinates between the $+/\times$ modes.
More straightforwardly, we find
\begin{align}
  \tilde{h}_+^{\rm O} \propto \tilde{\mathcal{A}} \,,
  \quad
  \tilde{h}_{\times}^{\rm O} \propto \tilde{\mathcal{B}} \,,
\end{align}
where we have used
\begin{align}
  C_+^2 &= (\mathcal{A}, \mathcal{A}) \cos^2 \dPhi + (\mathcal{B}, \mathcal{B}) \sin^2 \dPhi \,, \\
  C_{\times}^2 &= (\mathcal{B}, \mathcal{B}) \cos^2 \dPhi + (\mathcal{A}, \mathcal{A}) \sin^2 \dPhi \,, \\
  C_{\rm m}^2 &= - \left[ (\mathcal{A}, \mathcal{A}) - (\mathcal{B}, \mathcal{B}) \right] \cos \dPhi \sin \dPhi \,.
\end{align}
Therefore, the 2.5PN phase birefringence cannot be distinguished from the phase shift, and hence, we cannot constrain this type of modification.

\section{Constraining the CNCL model by taking into account both the amplitude and the phase birefringences}
\label{Sec:bothalphabeta}

\begin{figure}[tbp]
  \begin{center}
    \includegraphics[width=90mm]{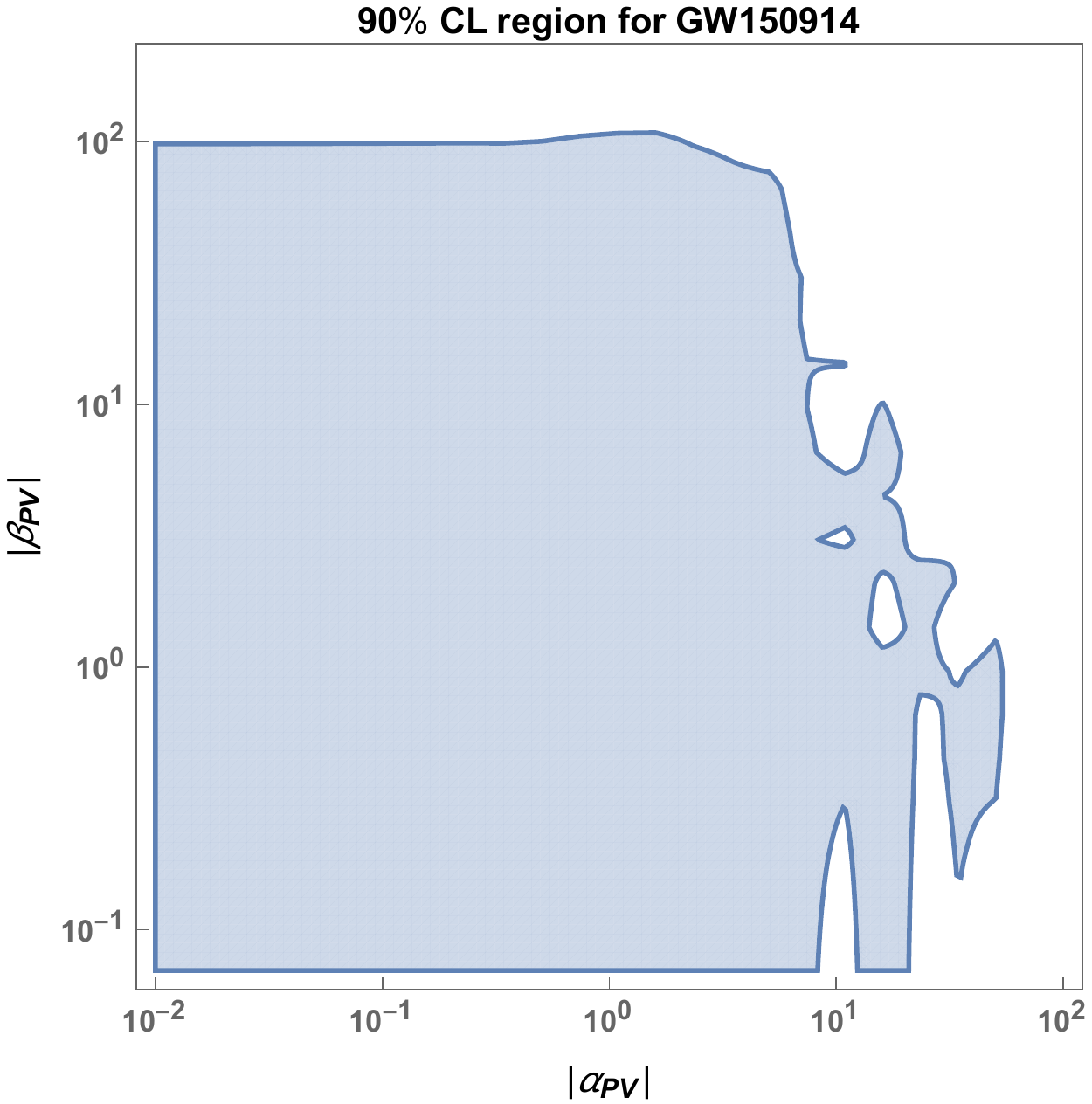}
  \end{center}
  \caption{The 90\% CL region on $\alpha_{\rm PV}$ and $\beta_{\rm PV}$.}
  \label{Fig:90CLalphabeta}
\end{figure}

In order to discuss a degeneracy between $\alpha_{\rm PV}$ and $\beta_{\rm PV}$, we perform a grid survey to find the best--fitting parameters by varying both $\alpha_{\rm PV}$ and $\beta_{\rm PV}$. 
To save the computational costs, in this analysis we fix the PN orders of the PV modifications to the 1.5PN amplitude and the 5.5PN phase birefringences, {\it i.e.} $(a_{\rm PV}, b_{\rm PV}) = (3, 6)$, which corresponds to the CNCL model. 
Figure~\ref{Fig:90CLalphabeta} shows the 90\% CL region for GW150914, which implies that at least in the CNCL model the degeneracy between $\alpha_{\rm PV}$ and $\beta_{\rm PV}$ is very weak. 
Similar results can be obtained from the other events. 
Therefore, we focus only on either the amplitude or the phase birefringences to obtain the constraints for various PN--order modifications.

\section{The secondary wave and noise on 4PN phase birefringence}
\label{Sec:4PNnoise}

\begin{figure}[tbp]
  \begin{center}
    \includegraphics[width=140mm]{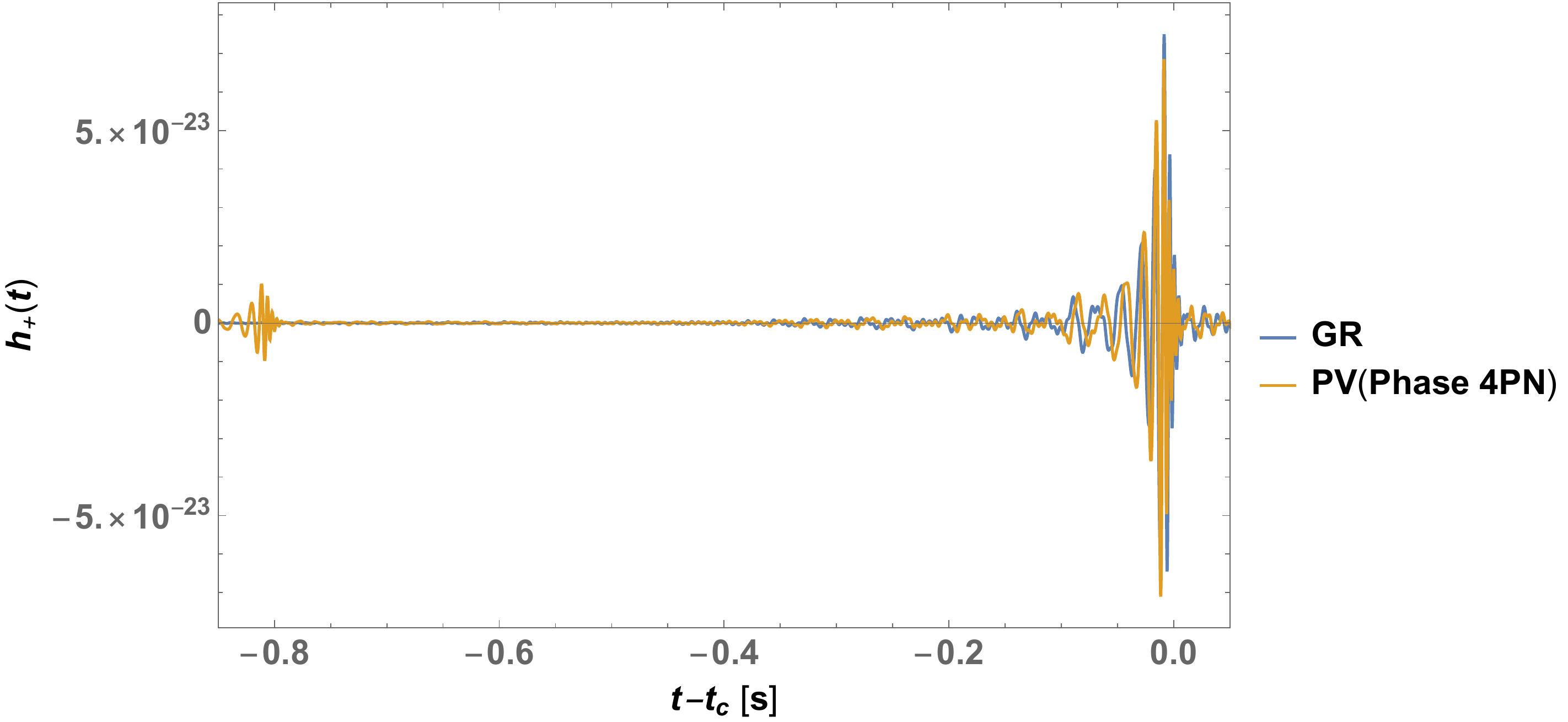}
  \end{center}
\caption{The whitened best--fitting templates for GW150914 in GR (blue) and in parity violation (orange) with the 4PN phase birefringence, in which the parameters are chosen as $(\iota, \beta) \approx (0.35 \pi, 5 \times 10^3)$. The coalescence time $t_c$ is chosen appropriately.}
\label{Fig:4PNbestfit}
\end{figure}

Basically, the 4PN phase birefringence corresponds to shifting the origin of time in the opposite directions for respective circular polarization modes. 
Therefore, $+/\times$--modes represented by this superposition can be discriminated from the GR template in principle. 
In this section, we consider nearly face--on binaries. 
In this case, it is difficult to constrain the parameter.
Figure~\ref{Fig:4PNbestfit} shows the best--fitting templates in GR (blue) and in PV gravity (orange) for the 4PN phase birefringence. 
Since the 4PN phase birefringence does not change the waveform of the dominant component, its contribution to the likelihoods is roughly the same as that in GR. 
Therefore, the difference is caused by the {\it ``secondary wave''} that comes before or after the primary one (around $-0.8 \, {\rm s}$ in Fig.~\ref{Fig:4PNbestfit}). 
Again, since the 4PN phase birefringence corresponds to the time--shift, the shape of the secondary wave is similar to the primary, but its amplitude depending on the inclination is not. 
Therefore, the match of the secondary wave with the data can be imitated by the match of the primary wave with noise around the signal in the data. 
The template is normalized in order for the standard deviation of the match with the Gaussian noise at unity. However, in practice, the standard deviation of the match can be much larger than unity because of the non--Gaussianity of the noise. This results in the contribution of the secondary wave to the likelihood becoming too large to constrain the deviation from GR. 
In order to constrain/probe the 4PN phase birefringence, therefore, the signals from nearly edge--on binaries, like Fig.~\ref{Fig:ExampleTemp}, is required.

\bibliographystyle{ptephy}
\bibliography{Ref}

\end{document}